\documentclass[10pt,letterpaper]{article}
\usepackage[utf8]{inputenc}
\usepackage{amsmath}
\usepackage{amsfonts}
\usepackage{amssymb}
\usepackage{authblk}
\usepackage{booktabs}
\usepackage{cite}
\usepackage{graphicx}
\usepackage[left=2cm,right=2cm,top=2cm,bottom=2cm]{geometry}
\usepackage{multirow}
\usepackage{subcaption}

\usepackage{color}

\author[1]{Pasquale Bosso\thanks{pasquale.bosso@uleth.ca}}
\author[1]{Saurya Das\thanks{saurya.das@uleth.ca}}
\author[2]{Robert B. Mann\thanks{rbmann@uwaterloo.ca}}
\affil[1]{Theoretical Physics Group and Quantum Alberta, University of Lethbridge,\protect\\ 4401 University Drive, Lethbridge, Alberta, Canada, T1K 3M4\vspace{1em}}
\affil[2]{Department of Physics and Astronomy, University of Waterloo,\protect \\ Waterloo, Ontario, Canada, N2L 3G1 and\protect \\Perimeter Institute, 31 Caroline St. N., Waterloo, Ontario, Canada, N2L 2Y5}

\title{Planck scale Corrections to the Harmonic Oscillator, Coherent and Squeezed States}

\date{}

\begin{document}

\maketitle

\begin{abstract}
	The Generalized Uncertainty Principle (GUP) is a modification of Heisenberg's Principle predicted by several theories of Quantum Gravity.
	It consists of a modified commutator between position and momentum.
	In this work we compute potentially observable effects that GUP implies for the harmonic oscillator, coherent and squeezed states in Quantum Mechanics.
	In particular, we rigorously analyze the GUP-perturbed harmonic oscillator Hamiltonian, defining new operators that act as ladder operators on the perturbed states.
	We use these operators to define the new coherent and squeezed states.
	We comment on potential applications.
\end{abstract}

\tableofcontents

\section{Introduction} 

One of the most active research areas in theoretical physics is the formulation of a quantum theory of gravity that would reproduce the well-tested theories of Quantum Mechanics (QM) and General Relativity (GR) at low energies.
The large energy scales necessary to test proposed theories of Quantum Gravity make this investigation extremely challenging.
Nonetheless, it is important to propose phenomenological models and tests of such theories at low energies, and the Generalized Uncertainty Principle (GUP) offers precisely this opportunity.

Many theories of QG suggest that there exists a momentum-dependent modification of  the Heisenberg Uncertainty Principle (HUP), and the consequent existence of a minimal measurable length \cite{Gross1988_1,Amati1989_1,Maggiore1993_1,Maggiore1993_2,Kempf1995_1,Garay1995_1,Scardigli1999_1,Capozziello2000,AmelinoCamelia2002_1}.
This modification is universal and affects every Hamiltonian, since it will affect the kinetic term.
In the last couple of decades several investigations have been conducted on many aspects and systems of QM, 
such as Landau levels \cite{Das2009_1,Ali2011_1}, Lamb shift, the case of a potential step and of a potential barrier
\cite{Das2009_1,Ali2011_1}, the case of a particle in a box \cite{Ali2009_1}, and the theory of angular momentum \cite{Bosso2016_1}.
Furthermore, potential experimental tests have been proposed considering microscopic \cite{Bawaj2014_1} or macroscopic Harmonic Oscillators (HO) \cite{Marin2013_1}, or using Quantum Optomechanics \cite{Pikovski2012_1,Bosso2016_2}.
Therefore our analysis is motivated by the fact that while very accurate systems with 
very little noise can be constructed, the energy perturbations and other results derived
in our paper will always be there, and potentially observable for highly 
sensitive systems. 
Similarly, any such deviations from the standard uncertainty profiles if observed, would
also be ascribed to new physics such as the GUP.

The most general modification of the HUP, was proposed in \cite{Ali2011_1} and includes linear and quadratic terms in momentum of the following form 
\begin{equation}
	[q,p] = i \hbar (1 - 2 \delta \gamma p + 4 \epsilon \gamma^2 p^2)  \label{eqn:GUP1}
\end{equation}
in 1 dimension, where $\delta$ and $\epsilon$ are two dimensionless parameters defining the particular model, and
\begin{equation}\label{Mpscale}
	\gamma = \frac{\gamma_0}{M_\mathrm{P} c}~,
\end{equation}
where $M_\mathrm{P}$ and $c$ are Planck mass and the speed of light, respectively, and $\gamma_0$ is a dimensionless parameter.
The presence of the quadratic term is dictated by string theory \cite{Gross1988_1,Amati1989_1} and \emph{gedanken} experiments in black hole physics \cite{Maggiore1993_1,Scardigli1999_1}, whereas the linear term is motivated by doubly special relativity \cite{AmelinoCamelia2002_1}, by the Jacobi identity of the corresponding $q_i,p_j$ commutators, and as a generalization of the quadratic model.
To incorporate both possibilities, we will leave the two parameters $\delta$ and $\epsilon$ undetermined, unless otherwise specified.
While it is possible to absorb $\gamma$ into  redefinitions of $\delta$ and $\epsilon$, it is in practice useful to keep $\gamma$ distinct so that it can function as an expansion parameter under various circumstances.

In the present work,  we revisit the problem of quantizing the HO \cite{Kempf1995_1} by incorporating the   GUP  \eqref{eqn:GUP1}, and rigorously investigate the implications of this for coherent and squeezed states.
Unlike previous investigations \cite{Bender1969,Dadic2003,Das2016,Quintela2016}, our focus is on a rigorous algebraic approach, which not only allows for a more efficient definition of the HO energy spectrum but also defines coherent and squeezed states of the HO with GUP \eqref{eqn:GUP1} in a consistent way.
Furthermore,   we consider for the first time  simultaneous perturbations in $p^3$ and $p^4$;  previous studies focused on a $p^4$ perturbation only.
We will also consider perturbation theory up to second order in $\gamma$, finding finite results up to that order.
We are thus able to derive potentially observable deviations from the standard HO due to GUP or Planck scale effects.
In particular, we find a model-dependent modification of the HO energy spectrum, as well as modified position and momentum uncertainties for coherent and squeezed states.
Similar analyses concerning coherent and squeezed states of the HO for noncommutative spaces were carried out in \cite{Dey2012,Dey2013,Dey2015}.
A similar problem was also addressed in \cite{Ghosh2012}, although a number of terms in the Hamiltonian arising from perturbation were neglected.

This paper is organized as follows.
In Sec. \ref{sec:HO_GUP}, we consider a GUP-induced perturbation of the HO, computing perturbed energy eigenstates and eigenvalues.
Moving beyond the previous treatment of the HO with the GUP \cite{Kempf1995_1}, the algebraic method adopted here turns out to be more concise, allowing for definitions that will be useful in the rest of the paper.
A new set of ladder operators for the perturbed states is defined in Sec. \ref{sec:new_operators}, and in Sec. \ref{sec:coherent_states}, we consider coherent states, focusing on their position and momentum uncertainties, and their time evolution.
A similar analysis for squeezed states for a HO is performed in Sec. \ref{sec:squeezed_states},  with special reference to specific GUP models   \cite{Kempf1995_1} and \cite{Ali2011_1}.
In Sec. \ref{sec:conclusions} we summarize our work, and comment on potential applications.

\section{Harmonic Oscillator in GUP} \label{sec:HO_GUP}

Consider the Hamiltonian for the one-dimensional HO
\begin{equation}
 H = \frac{p^2}{2m} + \frac{1}{2}m\omega^2 q^2 =
 -\frac{\hbar \omega}{4} (a^\dagger - a)^2 + \frac{\hbar\omega}{4}(a + a^\dagger)^2 = 
 \hbar\omega\left(N + \frac{1}{2}[a,a^\dagger]\right)~, \label{eqn:Hamiltonian}
\end{equation}
where $N = a^\dagger a$ is the usual number operator.
Though in the standard theory $[a,a^\dagger]=1$, it is easy to show that this relation is no longer valid once GUP is incorporated.

Using the model in (\ref{eqn:GUP1}) and expanding the physical position $q$ and momentum $p$ in terms of the low-energy quantities, $q_0$ and $p_0$ respectively, we obtain \cite{Das2009_1},
\begin{equation}
	q = q_0~, \qquad p = p_0 \left( 1 - \delta \gamma p_0 + 2 \gamma^2 \frac{\delta^2 + 2\epsilon}{3} p_0^2 \right) 
\end{equation}
where $[q_0,p_0] = i \hbar$.
In terms of these quantities, we can write the Hamiltonian for the harmonic oscillator as
\begin{equation}
	H = \frac{p_0^2}{2m} + \frac{1}{2}m\omega^2 q_0^2 - \delta \gamma \frac{p_0^3}{m} + \frac{7 \delta^2 + 8 \epsilon}{3} \gamma^2 \frac{p_0^4}{2 m}
	\label{eqn:perturbed_Hamiltonian}
\end{equation}
to leading order in the parameters.
Notice that we consider at the same time perturbations proportional to $p^3$ and $p^4$, in contrast to previous studies along these lines  \cite{Bender1969,Dadic2003,Das2016,Quintela2016}. 
Using perturbation theory with perturbation parameter $\gamma$,  we can define the state
\begin{equation}\label{Ksum}
|n^{(K)} \rangle = \sum_{k=0}^K \sigma_k \gamma^k |n_k \rangle
\end{equation}
to order $ \gamma^K$, where $ |n_k \rangle$ is the $k$-th order perturbation of the eigenvectors $ |n_0 \rangle =  |n  \rangle$ of
the unperturbed HO, whose eigenvalues are  $E^{(0)} = \hbar \omega (n + 1/2)$, and the $\sigma_k$ are constant coefficients
that are calculable from perturbation theory.  
It is tedious but straightforward to obtain the following expression for the 
normalized perturbed Hamiltonian eigenstates up to second order in $\gamma$
\begin{multline}
	|n^{(2)} \rangle = - \frac{\delta^2}{72} \gamma^2 \frac{\hbar m \omega}{2} \sqrt{n^{\underline{6}}} |n-6\rangle
	+ \frac{\gamma^2}{16} \frac{\hbar m \omega}{2} \left[ \delta^2 \left( 4n - \frac{2}{3} \right) + \frac{8 \epsilon}{3} \right] \sqrt{n^{\underline{4}}} |n-4\rangle
	- i \delta \frac{\gamma}{6} \sqrt{\frac{\hbar m \omega}{2}} \sqrt{n^{\underline{3}}}|n-3\rangle + \\
	- \frac{\gamma^2}{8} \frac{\hbar m \omega}{2} \left[ \delta^2 \left( 7n^2 - \frac{29}{3} n - \frac{11}{3} \right) + \frac{16 \epsilon}{3} (2n - 1) \right] \sqrt{n^{\underline{2}}} |n-2\rangle
	+ i \frac{3}{2} \delta \gamma \sqrt{\frac{\hbar m \omega}{2}} n \sqrt{n^{\underline{1}}} |n-1\rangle + \\
	+ \left[1 - \delta^2 \frac{\gamma^2}{72} \frac{\hbar m \omega}{2} (164 n^3 + 246 n^2 + 256 n + 87)\right]|n\rangle
	+ i \frac{3}{2} \delta \gamma \sqrt{\frac{\hbar m \omega}{2}} (n+1) \sqrt{(n+1)^{\overline{1}}} |n+1\rangle + \\
	- \frac{\gamma^2}{8} \frac{\hbar m \omega}{2} \left[ \delta^2 \left( 7n^2 + \frac{71}{3} n + 13 \right) - \frac{16 \epsilon}{3} (2n + 3) \right] \sqrt{(n+1)^{\overline{2}}} |n+2\rangle - i \delta \frac{\gamma}{6} \sqrt{\frac{\hbar m \omega}{2}} \sqrt{(n+1)^{\overline{3}}} |n+3\rangle + \\
	+ \frac{\gamma^2}{16} \frac{\hbar m \omega}{2} \left[ \delta^2 \left( 4n + \frac{14}{3} \right) - \frac{8 \epsilon}{3} \right] \sqrt{(n+4)^{\overline{4}}} |n+4\rangle
	- \frac{\delta^2}{72} \gamma^2\frac{\hbar m \omega}{2} \sqrt{(n+1)^{\overline{6}}} |n+6\rangle~, 
	\label{eqn:perturbed_state}
\end{multline}
where we used the notation
\begin{align}
	(n+1)^{\overline{k}} & = \frac{(n+k)!}{n!} ~, & n^{\underline{k}} & = \frac{n!}{(n-k)!} \quad \mbox{for }k \leq n~, & n^{\underline{k}} & = 0 \quad \mbox{for } k > n ~.
\end{align}
The perturbed energy eigenvalue is
\begin{equation}
	E^{(2)} = \hbar \omega \left\{ \left( n + \frac{1}{2} \right) - \frac{\hbar m \omega}{2} \gamma^2 [ (4 n^2 + 4 n + 1) \delta^2 - (2 n^2 + 2 n + 1) 2 \epsilon ] \right\} = E^{(0)} + \Delta E ~, \label{eqn:energ_eigenvalue}
\end{equation}
where
\begin{align}
	\Delta E = - \hbar \omega \frac{\hbar m \omega}{2} \gamma^2 [ (4 n^2 + 4 n + 1) \delta^2 - (2 n^2 + 2 n + 1) 2 \epsilon]~.
\end{align}
Furthermore, the spacing of the energy levels is
\begin{equation}
	E^{(2)}(n+1) - E^{(2)}(n) = \hbar \omega - 8 \hbar \omega (n + 1) \frac{\hbar m \omega}{2} \gamma^2 [ \delta^2 - \epsilon ]~.
\end{equation}
Note that the linear and the quadratic contributions to \eqref{eqn:GUP1}, identifiable through the parameters $\delta$ and $\epsilon$, make opposing contributions to the energy. 
It is therefore worth noting that, for the class of models with $\delta^2 = \epsilon$, the correction $\Delta E$ is independent of $n$, resulting in an equally spaced energy spectrum.
For $\delta^2 < \epsilon$, the correction is a positive function of the number $n$.
In particular, for $\delta = 0$ and $\epsilon = 1/4$, we obtain the results presented in \cite{Kempf1995_1} and \cite{Hossenfelder2003}.
Finally, we see that for $\delta^2 > \epsilon$ the spacing between energy levels decreases and we find the value of $n$ corresponding to the maximal energy
\begin{equation}
	n_\mathrm{max} = \left\lfloor \frac{1}{8 \frac{\hbar m \omega}{2} \gamma^2 (\delta^2 - \epsilon)} - 1 \right\rfloor~.
\end{equation}
Notice that this depends on the Planck scale parameters, and the energy corresponding to the above $n$ signals the breakdown of the GUP model, and necessitates higher order terms.

To estimate the magnitude of these corrections, we compute
\begin{equation}
	\frac{|\Delta E|}{E^{(0)}} = \frac{\hbar m \omega}{2} \gamma^2 \frac{\left|(4 n^2 + 4 n + 1) \delta^2 - (2 n^2 + 2 n + 1) 2 \epsilon\right|}{n + \frac{1}{2}} \stackrel{n\gg1}{\simeq} 4 n \frac{\hbar m \omega}{2} \gamma^2 \left|\delta^2 - \epsilon\right| \simeq 2 m E^{(0)} \gamma^2 \left|\delta^2 - \epsilon\right|~,
\end{equation}
Conversely, if we performed an experiment with a sensitivity $\Delta = |\Delta E|/ E^{(0)}$, we could detect a deviation from the unperturbed energies at
\begin{align}
&	E^{(0)} =   \frac{\Delta}{2 m \gamma^2 \left|\delta^2 - \epsilon\right|} = \frac{21.279 \Delta}{m \gamma_0^2 \left|\delta^2 - \epsilon\right|} \mathrm{J}  \qquad \textrm{or} \quad
	n =   \frac{\Delta}{4 \frac{\hbar m \omega}{2} \gamma^2 \left|\delta^2 - \epsilon\right|} = \frac{20.169 \Delta}{ m \omega \gamma_0^2 \left|\delta^2 - \epsilon\right|} \times 10^{34}
	 \label{eqn:detection}
\end{align}
for $m$ and $\omega$ measured in Kg and in Hz, respectively.
Although this value  appears huge, as we show below   these Planck scale effects are potentially accessible to a number of current and future experiments.
We in particular see that massive and/or rapidly oscillating systems most significantly enhance GUP effects.

As particular examples, we considered the cases of the mechanical oscillator considered in \cite{Pikovski2012_1}, the oscillators considered in \cite{Bawaj2014_1}, the resonant-mass bar AURIGA in its first longitudinal mode \cite{Marin2013_1}, and the mirrors in LIGO oscillating at a frequency in the middle of its detection band \cite{Abbott2016}.
\begin{table}
\begin{center}
\begin{tabular}{lcccc}
	\toprule
	Type & Ref. & $m$ (Kg) & $\omega/2\pi$ (Hz) & $n / \displaystyle{\frac{\Delta}{\gamma_0^2 \left|\delta^2 - \epsilon\right|}}$ \\
	\midrule
	Optomechanical system & \cite{Pikovski2012_1} & $10^{-11}$ & $10^5$ & $ 3 \times 10^{40}$ \\
	Bar detector AURIGA &\cite{Marin2013_1} & $1.1 \times 10^{3}$ & $900$ & $3 \times 10^{28}$ \\
	\multirow{4}{*}{Mechanical oscillators} & \multirow{4}{*}{\cite{Bawaj2014_1}} & $3.3 \times 10^{-5}$ & $5.64 \times 10^3$ & $2 \times 10^{35}$ \\
	& & $7.7 \times 10^{-8}$ & $1.29 \times 10^5$ & $3 \times 10^{36}$ \\
	& & $2 \times 10^{-8}$ & $1.42 \times 10^5$ & $10^{37}$ \\
	& & $2 \times 10^{-11}$ & $7.47 \times 10^5$ & $2 \times 10^{39}$ \\
	LIGO detector & \cite{Abbott2016} & 40 & 200 & $4 \times 10^{30}$\\
	\bottomrule
\end{tabular}
%\begin{tabular}{l|c|c|c|}
%	\cline{2-4}
%	& $m$ (Kg) & $\omega/2\pi$ (Hz) & $n / \displaystyle{\frac{\Delta}{\gamma_0^2 \left|\delta^2 - \epsilon\right|}}$ \\
%	\hline
%	\multicolumn{1}{|l|}{Optomechanical system \cite{Pikovski2012_1}} & $10^{-11}$ & $10^5$ & $ 3 \times 10^{40}$ \\
%	\multicolumn{1}{|l|}{Bar detector AURIGA \cite{Marin2013_1}} & $1.1 \times 10^{3}$ & $900$ & $3 \times 10^{28}$ \\
%	\multicolumn{1}{|l|}{\multirow{4}{*}{Mechanical oscillators \cite{Bawaj2014_1}$ \left\{ \begin{array}{c} ~ \\ ~ \\ ~ \\ ~ \end{array} \right.$}} & $3.3 \times 10^{-5}$ & $5.64 \times 10^3$ & $2 \times 10^{35}$ \\
%	\multicolumn{1}{|l|}{}& $7.7 \times 10^{-8}$ & $1.29 \times 10^5$ & $3 \times 10^{36}$ \\
%	\multicolumn{1}{|l|}{}& $2 \times 10^{-8}$ & $1.42 \times 10^5$ & $10^{37}$ \\
%	\multicolumn{1}{|l|}{}& $2 \times 10^{-11}$ & $7.47 \times 10^5$ & $2 \times 10^{39}$ \\
%	\multicolumn{1}{|l|}{LIGO detector \cite{Abbott2016}} & 40 & 200 & $4 \times 10^{30}$\\
%	\multicolumn{1}{|l|}{GW150914 \cite{Abbott2016}} & $3.2 \times 10^{31}$ & 100 & 10\\
%	\hline
%\end{tabular}
\caption{Some relevant examples of HO are considered, including their mass, frequency and levels at which GUP effects become dominant, as given by \eqref{eqn:detection}.} \label{tbl:detection}
\end{center}
\end{table}

\section{New GUP modified ladder operators} \label{sec:new_operators}

Given the form of the perturbed eigenstates in terms of the standard number states (\ref{eqn:perturbed_state}), it is easy to show that the standard annihilation and creation operators do not act anymore as ladder operators.
We therefore define a new set of operators, useful for constructing coherent and squeezed states, such that
\begin{align}
	\tilde{a} | n^{(2)} \rangle = & \sqrt{n} ~ | (n-1)^{(2)} \rangle~, & \widetilde{a^\dagger} | n^{(2)} \rangle = & \sqrt{n+1} ~ | (n+1)^{(2)} \rangle~, & \tilde{N} | n^{(2)} \rangle = & n ~ | n^{(2)} \rangle ~.
\label{defs:tilde_ops}	
\end{align}
Note that these operators obey the following relations
\begin{align}
	\widetilde{a^\dagger} = & \tilde{a}^\dagger~, & \tilde{N} = & \tilde{a}^\dagger \tilde{a} ~, & [\tilde{a},\tilde{a}^\dagger] = & 1~.
\end{align}
Using these definitions, we find the following expressions for the  operators in \eqref{defs:tilde_ops} in terms of the standard annihilation, creation and number operators:
\begin{subequations} \label{eqns:new_ops}
\begin{align}
	\tilde{a} = & a
	- \delta i \frac{\gamma}{2} \sqrt{\frac{\hbar m \omega}{2}} \left[ 3 a^2 + 3 (2N+1) - {a^\dagger}^2 \right] + \nonumber \\
	& - 2 \gamma^2 \frac{\hbar m \omega}{2} \left[ \left( \frac{5}{3} \delta^2 - \frac{2}{3} \epsilon \right) a^3 + \delta^2 a N + \left( \delta^2 + 2 \epsilon \right) N a^\dagger - \left( \frac{2}{3} \delta^2 + \frac{1}{3} \epsilon \right) {a^\dagger}^3 \right]~, 
\label{a-tilde}\\
	\widetilde{a^\dagger} = & a^\dagger 
	+ \delta i \frac{\gamma}{2} \sqrt{\frac{\hbar m \omega}{2}} \left[ 3{a^\dagger}^2 + 3(2N+1) - a^2 \right] + \nonumber \\
	& - 2 \gamma^2 \frac{\hbar m \omega}{2} \left[ \left( \frac{5}{3} \delta^2 - \frac{2}{3} \epsilon \right) {a^\dagger}^3 + \delta^2 N a^\dagger + ( \delta^2 + 2 \epsilon ) a N - \left(\frac{2}{3} \delta^2 + \frac{1}{3} \epsilon \right) a^3 \right]~, 
	\label{a-tilde-dag} \displaybreak\\
	\tilde{N} = & N 
	- \delta i \frac{\gamma}{2} \sqrt{\frac{\hbar m \omega}{2}} \left[ a^3 - 3 ( a N - N a^\dagger ) - {a^\dagger}^3 \right] + \nonumber \\
	& + \frac{\gamma^2}{2} \frac{\hbar m \omega}{2} \left\{ \frac{7 \delta^2 + 8 \epsilon}{6} [ a^4 - 2 (2 a N a + a^2) - 2 (2 a^\dagger N a^\dagger + {a^\dagger}^2) + {a^\dagger}^4 ] + \frac{\delta^2}{2} (30 N^2 + 30 N + 11) \right\} 
\label{N-tilde}	
\end{align}
\end{subequations}
To obtain these relations, we used the following procedure.
Consider, for example, the annihilation operator $\tilde{a}$ defined by (\ref{defs:tilde_ops}).
Applying this operator on the perturbed number state (\ref{eqn:perturbed_state}), we notice that to the 0-th order in $\gamma$ we get $\tilde{a} |n\rangle = a |n\rangle$, as expected. Requiring \eqref{defs:tilde_ops} to hold to order $\gamma$, 
we compute 
\begin{equation}
	(\tilde{a} - a) \left|n^{(1)}\right\rangle = \sqrt{n} \left|(n-1)^{(1)} \right> - a \left|n^{(1)} \right> = - \delta i \frac{\gamma}{2} \sqrt{\frac{\hbar m \omega}{2}} \left[ 3 a^2 + 3 (2N+1) - {a^\dagger}^2 \right] \left|n^{(1)} \right> 
\end{equation}
where $\left|n^{(1)}\right\rangle$ is obtained  from the order-$\gamma$ terms in \eqref{eqn:perturbed_state}.	
Therefore $\tilde{a} = a - \delta i \frac{\gamma}{2} \sqrt{\frac{\hbar m \omega}{2}} \left[ 3 a^2 + 3 (2N+1) - {a^\dagger}^2 \right]$
 to first order in $\gamma$.  Repeating this procedure to order $\gamma^2$ yields 
\begin{multline}
	\left\{\tilde{a} - a + \delta i \frac{\gamma}{2} \sqrt{\frac{\hbar m \omega}{2}} \left[ 3 a^2 + 3 (2N+1) - {a^\dagger}^2 \right] \right\} \left| n^{(2)} \right> = \\
	- 2 \gamma^2 \frac{\hbar m \omega}{2} \left[ \left( \frac{5}{3} \delta^2 - \frac{2}{3} \epsilon \right) a^3 + \delta^2 a N + \left( \delta^2 + 2 \epsilon \right) N a^\dagger - \left( \frac{2}{3} \delta^2 + \frac{1}{3} \epsilon \right) {a^\dagger}^3 \right] \left| n^{(2)} \right> 
\end{multline}
which gives \eqref{a-tilde}.
A similar procedure can be used for $\tilde{a^\dagger}$ and $\tilde{N}$.
Continuing this procedure, one can extend the relations in (\ref{eqns:new_ops}) to any arbitrary order.

Using these expressions, the Hamiltonian of an harmonic oscillator with GUP can be written in a simple form:
\begin{equation}
	H = \hbar \omega \left\{ \left( \tilde{N} + \frac{1}{2} \right) - \frac{\hbar m \omega}{2} \gamma^2 [ 4 (\tilde{N}^2 + \tilde{N}) (\delta^2 - \epsilon) + \delta^2 - 2 \epsilon] \right\} 
	\label{eqn:Hamiltonian_tilde}
\end{equation}
as well as  
\begin{subequations} \label{eqn:expansions_q&p}
\begin{align}
	q = & (\tilde{a}^\dagger + \tilde{a}) \sqrt{\frac{\hbar}{2 m \omega}} 
		- 2 i (\tilde{a}^\dagger {}^2 - \tilde{a}^2 ) \delta \frac{\hbar}{2} \gamma 
		- 2 [ ( \tilde{a}^3 + \tilde{a}^\dagger {}^3 ) ( 2 \delta^2 + \epsilon) + (\tilde{a} \tilde{N} + \tilde{N} \tilde{a}^\dagger) (3 \delta^2 - 2 \epsilon) ] \frac{\hbar}{2} \sqrt{\frac{\hbar m \omega}{2}} \gamma^2 \\
 	p = & i (\tilde{a}^\dagger - \tilde{a}) \sqrt{\frac{\hbar m \omega}{2}} 
 		+ 2 (\tilde{a}^\dagger {}^2 + 2 \tilde{N} + 1 + \tilde{a}^2) \delta \frac{\hbar m \omega}{2} \gamma 
 		+ 2 i [( \tilde{a}^3 - \tilde{a}^\dagger {}^3 ) ( 2 \delta^2 + \epsilon) + 3 (\tilde{a} \tilde{N} - \tilde{N} \tilde{a}^\dagger ) \delta^2 ] \left(\frac{\hbar m \omega}{2}\right)^{3/2} \gamma^2 
\end{align}
\end{subequations}
for the physical position and momentum operators.
It is then easy to find the expectation values and the uncertainties in position and momentum for the perturbed number states.
Defining $ \langle A \rangle = \langle  n^{(2)}| A | n^{(2)} \rangle$ for any operator $A$, we find
\begin{subequations}
\begin{align}
	\langle q \rangle = & 0 ~, \\
	\langle q^2 \rangle = & \frac{\hbar}{2 m \omega} (2 n + 1) - 4 \gamma^2 \frac{\hbar^2}{4} \left[ \delta^2 (2n + 1)^2 - 2 \epsilon (2 n^2 + 2 n + 1) \right]~ ,\\
	(\Delta q)^2 = & \langle q^2 \rangle ~, \\
	\langle p \rangle = & 2 \delta \gamma \frac{\hbar m \omega}{2} (2n + 1) ~,  \label{19d}	 \\
	\langle p^2 \rangle = & \frac{\hbar m \omega}{2} (2n + 1)~,\\
	(\Delta p)^2 = & \frac{\hbar m \omega}{2} (2n + 1) \left[ 1 - 4 \delta^2 \gamma^2 \frac{\hbar m \omega}{2} (2n + 1) \right]~, \label{19f}\\
	(\Delta q)^2 (\Delta p)^2 = & \frac{\hbar^2}{4} (2n + 1)^2 - 8 \gamma^2 \frac{\hbar^2}{4} \frac{\hbar m \omega}{2} (2n + 1) [ \delta^2 (2n + 1)^2 - \epsilon (2n^2 + 2n + 1)]
\end{align}
\end{subequations}

The presence of $\delta$ in \eqref{19d} and \eqref{19f} leads to interesting features.
First, the expectation value of the momentum does not vanish whereas the expectation value of the position does.
This is a consequence of the linear part of the GUP \eqref{eqn:GUP1}, which singles out a preferred direction and so breaks translation invariance.
Furthermore, this same term implies that  there exists a value for $n$ such that $(\Delta p)^2 < 0$, showing that a linear model leads to critical results for high energy systems, \emph{i.e.} when the energy is close to the Planck scale.
We expect that at this energy scale a full theory of Quantum Gravity has to be considered, with non-negligible higher order corrections.

To conclude this section,  we observe that the definitions in \eqref{defs:tilde_ops} and the methods developed in this section can be easily extended to any perturbation 
that is of the form of a polynomial of $q$ and $p$ {\it and} to any order in perturbation theory.
Using this method, the full Hamiltonian can be written in the following form
\begin{equation}
	H = \hbar \omega \left\{ \left(\tilde{N} + \frac{1}{2} \right) + \sum_{i=0}^{S} c_i \tilde{N}^i \right\}~, \qquad \mbox{with} \qquad S = \left\{ \begin{array}{cc} 0 \quad & \mbox{for } d \mbox{ an odd number and } e = 1\\
    \lfloor (d + e - 1)/2 \rfloor \quad & \mbox{otherwise}
    \end{array} \right. \label{eqn:gen_Hamiltonian}
\end{equation}
where $d$ is the degree of the polynomial representing the perturbation and $e$ the order of the perturbation theory.
The coefficients $c_i$ are constants that depend on the parameters of the perturbation.
In this paper we consider the case with
\begin{subequations}
\begin{align}
	d = & 4~, & e = & 2~, & S = 2~,\\
    c_0 = & -\frac{\hbar m \omega}{2}\gamma^2 (\delta^2 - 2\epsilon)~, & c_1 = & c_2 = - 4 \frac{\hbar m \omega}{2}\gamma^2 (\delta^2 - \epsilon)~.
\end{align}
\end{subequations}
Furthermore, we emphasize that $|n^{(2)}\rangle$ are the eigenstates of the full Hamiltonian with GUP (\ref{eqn:Hamiltonian}) up to $\mathcal{O}(\gamma^2)$, while the standard states $|n\rangle$ are not.
Therefore, we will consider the former as the physical number states.
Furthermore, the $\sim$-operators that we have defined have the same properties, in terms of commutators and actions on number states, as the analogous operators of the standard theory.
For these reasons, we apply the above new operators to define coherent and squeezed states.

\section{Coherent States} \label{sec:coherent_states}

As in the standard theory, we define a coherent state as the eigenstate of the annihilation operator $\tilde{a}$
\begin{equation}
	\tilde{a} | \alpha^{(2)} \rangle = \alpha |\alpha^{(2)} \rangle  \label{def:coherent_states}
\end{equation}
to order $\gamma^2$.
This is because $|n^{(2)}\rangle$ is the physical number state and $\tilde{a}$ is its operator.
Following the results of the previous section, we then have
\begin{equation}
	|\alpha^{(2)} \rangle = e^{- \frac{|\alpha|^2}{2}} \sum_{n=0}^\infty \frac{\alpha^n}{\sqrt{n!}} |n^{(2)}\rangle~, \label{eqn:expansion_coherent_states}
\end{equation}
\emph{i.e.} a Poisson distribution of number states in $\sim$-coherent states\footnote{We can also define a displacement operator of the form $\displaystyle{	\tilde{\mathcal{D}}(\alpha) = e^{\alpha \tilde{a}^\dagger - \alpha^* \tilde{a}}}$.}.
Furthermore, given the expansions in (\ref{eqn:expansions_q&p}), we can easily obtain the following results for the uncertainties
\begin{subequations} \label{eqns:uncs_coherent}
\begin{align}
	\langle q \rangle = & \left( \alpha^\star + \alpha \right) \sqrt{\frac{\hbar}{2 m \omega}} 
	- 2 i \delta \left( {\alpha^\star}^2 - \alpha^2\right) \frac{\hbar}{2} \gamma + \nonumber \\
	& + \left[ (2 \epsilon - 3 \delta^2) (\alpha^\star + \alpha) ( 1 + |\alpha|^2) - ( \epsilon + 2 \delta^2) \left( {\alpha^\star}^3 + \alpha^3 \right) \right] \frac{\hbar}{2} \sqrt{\frac{\hbar m \omega}{2}} \gamma^2 ~, \\
	(\Delta q)^2 = & \frac{\hbar}{2 m \omega} - 4 i \delta \left( \alpha^\star - \alpha\right) \frac{\hbar}{2} \sqrt{\frac{\hbar}{2 m \omega}} \gamma + \nonumber \\
	& - 2 \left[ \epsilon \left(- 4 + \alpha^2 - 8 |\alpha|^2 + {\alpha^\star}^2\right) + \delta^2 \left( 2 + 9 \alpha^2 + 4 |\alpha|^2 + 9 {\alpha^\star}^2 \right) \right] \frac{\hbar^2}{4} \gamma^2~,\\
	\langle p \rangle = & i \left(\alpha^\star-\alpha\right) \sqrt{\frac{\hbar m\omega}{2}} + 2 \delta \left( {\alpha^\star}^2 + 2 |\alpha|^2 + \alpha^2 + 1 \right) \frac{\hbar m \omega}{2} \gamma + \nonumber \\
	& - i \left[ 3 \delta^2 (1 + |\alpha|^2) (\alpha^\star - \alpha) + (\epsilon + 2 \delta^2) \left( {\alpha^\star}^3 - \alpha^3\right) \right] \left( \frac{\hbar m \omega}{2} \right)^{3/2} \gamma^2~, \\
	(\Delta p)^2 = & \frac{\hbar m \omega}{2} + 2 \left(\frac{\hbar m \omega}{2}\right)^2 \gamma^2 \left[ ( {\alpha^\star}^2 + \alpha^2) (5 \delta^2 - 3 \epsilon) + (4 |\alpha|^2 - 2)\delta^2 \right] ~,\displaybreak\\
(\Delta q)^2 (\Delta p)^2 = & \frac{\hbar^2}{4} \left\{ 1
		- 4 i \delta \left( \alpha^\star - \alpha \right) \sqrt{\frac{\hbar m \omega}{2}} \gamma 
		+8 \left[ \epsilon \left( 1 + \alpha - \alpha^\star \right) \left( 1 - \alpha + \alpha^\star \right) - \delta^2 \left( 1 + \alpha^2 + {\alpha^\star}^2 \right) \right] \frac{\hbar m \omega}{2} \gamma^2 \right\}
\label{eqn:prod_unc}
\end{align}
\end{subequations}
where we see that the last term exhibits interesting properties.
If $\epsilon > 0$ and $\delta=0$ then the uncertainty $(\Delta q)^2 (\Delta p)^2 $ is greater than $\frac{\hbar^2}{4}$ whereas if $\epsilon = 0$ and $\delta>0$ then this quantity is smaller than $\frac{\hbar^2}{4}$.

The same situation holds for the   minimal uncertainty product, which is the smallest uncertainty product predicted by QM as found using the Schr\"odinger-Robertson relation.  
For the model in (\ref{eqn:GUP1}), we obtain 
\begin{multline}
	[(\Delta q)^2 (\Delta p)^2]_\mathrm{min} =  \left| \frac{\langle qp + pq \rangle}{2} - \langle q \rangle \langle p \rangle \right|^2 +  \frac{|\langle[q,p]\rangle|^2}{4} = \\
	= \frac{\hbar^2}{4} \left\{ 1
		- 4 i \delta \left( \alpha^\star - \alpha \right) \sqrt{\frac{\hbar m \omega}{2}} \gamma 
		+8 \left[ \epsilon \left( 1 + \alpha - \alpha^\star \right) \left( 1 - \alpha + \alpha^\star \right) - \delta^2 \left( 1 + \alpha^2 + {\alpha^\star}^2 \right) \right] \frac{\hbar m \omega}{2} \gamma^2 \right\}
\end{multline}
 and again nonzero $\epsilon$ 
makes this quantity larger than the standard QM value whereas non-zero $\delta$ makes it smaller.

Finally we note that   
\begin{equation}
	(\Delta q)^2 (\Delta p)^2 - [(\Delta q)^2 (\Delta p)^2]_\mathrm{min} =  0 \qquad \forall \alpha, \gamma, \delta, \epsilon~ . 
	\label{eqn:diff_coherent}
\end{equation}
and so coherent states are also minimal uncertainty states   when the GUP is included.
The quantity $(\Delta q)^2 (\Delta p)^2$ is the actual product of the uncertainties in position and momentum.
 Therefore, we see that coherent states as defined in \eqref{def:coherent_states} satisfy all the known properties of the standard theory, \emph{i.e.} they follow a Poisson distribution, they are eigenstates of the annihilation operator, and are minimal uncertainty states.
 
The uncertainty product $(\Delta q)^2 (\Delta p)^2$ in \eqref{eqn:prod_unc} is negative for some values of $\alpha$ and of $\sqrt{\frac{\hbar m \omega}{2}} \gamma$. This is due to the linear term in GUP and will go negative for
\begin{equation}
	\delta^2 > \frac{1 + 4 \Im^2(\alpha)}{1 + 2 \Re^2 (\alpha)} \epsilon~.
\end{equation}
In this case, the model will present anomalies when the characteristic momentum of the system fulfills the following relation
\begin{equation}
	\sqrt{\frac{\hbar m \omega}{2}} \gamma = \frac{ - \Im (\alpha) \delta \pm \sqrt{ 2 \left[1 + 2 \Re^2 (\alpha) \right] \delta^2 - 2 \left[1 + 4 \Im^2 (\alpha) \right] \epsilon}}{ 2 \left[1 + 2 \Re (\alpha^2) \right] \delta^2 - 2 \left[ 1 + 4 \Im^2 (\alpha) \right] \epsilon}~. \label{eqn:negative_unc_pro}
\end{equation}
Notice that this quantity depends on the phase of $\alpha$ and that one cannot \emph{a priori} choose the sign in the expression.
 
\subsection{Time Evolution} \label{subsec:time_coherent}

To consider the time-evolved coherent states, we analyze them in Heisenberg picture.
Since we seek to understand the evolution of the variances \eqref{eqns:uncs_coherent}, we first write the time-evolved annihilation operator 
\begin{equation}\label{a-tilde-t}
	\tilde{a}(t) = \tilde{a}(0) \exp \left\{ - i \omega t \left[ 1 - 8 \tilde{N} \left( \delta^2 - \epsilon \right) \frac{\hbar m \omega}{2} \gamma^2 \right] \right\} \equiv \tilde{a}(0) \exp \left[ - i \omega t \left( 1 - 8 \tilde{N} \chi \right) \right] ~,
\end{equation}
where we have defined the dimensionless quantity
\begin{equation}
	\chi = \left( \delta^2 - \epsilon \right) \frac{\hbar m \omega}{2} \gamma^2~.
\end{equation}
Notice that for the number operator we have $\tilde{N}(t) = \tilde{N}(0)$, and therefore the statistical properties of the coherent states do not change.

Using the time-evolved operators \eqref{a-tilde-t}, we find 
\begin{subequations}
\begin{align}
	\left( \Delta q \right)^2 = & \frac{\hbar}{2 m \omega} \left\{ 1 + 2 |\alpha|^2 \left[ 1 - e^{- 4 |\alpha|^2 \sin^2 \left[4 \chi \omega t\right]} \right] \right\}  \nonumber \\
	& + \alpha^2 \frac{\hbar}{2 m \omega} \left\{ e^{-|\alpha|^2 \left[1 - e^{16 i \chi \omega t}\right] - 2 i \omega t \left[1 - 12 \chi \right]} 
	-  e^{- 2 |\alpha|^2 \left[1 - e^{8 i \chi \omega t} \right] - 2 i \omega t \left[1 - 8 \chi \right] } \right\}   \nonumber \\
	& + \alpha^\star{}^2 \frac{\hbar}{2 m \omega} \left\{ e^{-|\alpha|^2 \left[1 - e^{-16 i \chi \omega t}\right] + 2 i \omega t \left[1 - 12 \chi\right]} 
	-  e^{-2 |\alpha|^2 \left[1 - e^{- 8 i \chi \omega t} \right]+ 2 i \omega t \left[1 - 8 \chi \right] } \right\}  \nonumber \\
	& + 4 i \sqrt{\frac{\hbar}{2 m \omega}} \frac{\hbar}{2} \left( \alpha e^{- i \omega t } - \alpha^\star e^{i \omega t} \right) \delta \gamma
	- 2 \frac{\hbar^2}{4}\left[ 2 (\delta^2 - 2 \epsilon) ( 1 - 2 |\alpha|^2) + (9 \delta^2 + \epsilon) \left( \alpha^2 e^{- 2 i \omega t} + \alpha^\star {}^2 e^{2 i \omega t} \right) \right] \gamma^2 ~, \\
	\left( \Delta p \right)^2 = & \frac{\hbar m \omega}{2}\left\{1 + 2|\alpha|^2\left[1 - e^ {- 4 |\alpha|^2 \sin^2 \left[4 \chi \omega t\right] } \right]\right\}   \nonumber \\
	& + \alpha^2 \frac{\hbar m \omega}{2} \left\{ e^{-2 |\alpha|^2 \left[1 - e^{8 i \chi \omega t} \right] - 2 i \omega t \left[1 - 8 \chi \right]} 
	- e^{ - |\alpha|^2 \left[1 - e^{16 i \chi \omega t }\right] - 2 i \omega t \left[1 - 12 \chi \right]} \right\}   \nonumber \\
	& + \alpha^\star {}^2 \frac{\hbar m \omega}{2} \left\{ e^{- 2 |\alpha|^2 \left[1 - e^{- 8 i \chi \omega t } \right] + 2 i \omega t \left[1 - 8 \chi t \right]} - e^{- |\alpha|^2 \left[1 - e^{- 16 i \chi \omega t } \right] + 2 i \omega t \left[1 - 12 \chi t \right] } \right\}   \nonumber \\
	& + 2 \left(\frac{\hbar m \omega}{2} \right)^2 \left[ \left(5 \delta^2-3 \epsilon\right) \left(\alpha^2 e^ {- 2 i \omega t } + \alpha^\star {}^2 e^{2 i \omega t}\right) - 2 \delta^2 (1 - 2 |\alpha|^2) \right] \gamma^2 
	\end{align}
\end{subequations}
for the uncertainties in position and momentum.

These two expressions indicate that the GUP imputes important effects on the uncertainties of position and momentum.
Consider the first line of each expression.
We notice an oscillation in the uncertainties, governed by the term $\sin^2 \left[4 \chi \omega t\right]$, with amplitude proportional to the standard uncertainties multiplied by $|\alpha|^2$.
This term, initially 0 for $t=0$, varies with time with a period
\begin{equation}
	T = \left| \frac{\pi}{4 \chi \omega} \right|~. \label{eqn:unc_period}
\end{equation}
 It is worth noticing that the period depends on $m$,  $\omega$, and the GUP parameters through the quantity $\chi$.
The subsequent two terms in each expression present an oscillation with a similar period.
This oscillation is present only for models with $\delta^2 \not = \epsilon$.
For the systems considered at the end of Sec. \ref{sec:HO_GUP}, we have the Table \ref{tbl:unc_osc_period}.
\begin{table}
\begin{center}
\begin{tabular}{lcccc}
\toprule
	Type & Ref. & $m$ (Kg) & $\omega/2\pi$ (Hz) & $T \gamma_0^2 \left|\delta^2 - \epsilon\right| (s)$ \\
	\midrule
	Optomechanical system & \cite{Pikovski2012_1} & $10^{-11}$ & $10^5$ & $ 1.61 \times 10^{35}$ \\
	Bar detector AURIGA & \cite{Marin2013_1} & $1.1 \times 10^{3}$ & $900$ & $1.81 \times 10^{25}$ \\
	\multirow{4}{*}{Mechanical oscillators} & \multirow{4}{*}{\cite{Bawaj2014_1}} & $3.3 \times 10^{-5}$ & $5.64 \times 10^3$ & $1.53 \times 10^{31}$ \\
	&& $7.7 \times 10^{-8}$ & $1.29 \times 10^5$ & $1.26 \times 10^{31}$ \\
	& & $2 \times 10^{-8}$ & $1.42 \times 10^5$ & $3.99 \times 10^{31}$ \\
	& & $2 \times 10^{-11}$ & $7.47 \times 10^5$ & $1.44 \times 10^{33}$ \\
	LIGO detector &\cite{Abbott2016} & 40 & 200 & $1.00 \times 10^{28}$\\
	\bottomrule
%	\cline{2-4}
%	& $m$ (Kg) & $\omega/2\pi$ (Hz) & $T \gamma_0^2 \left|\delta^2 - \epsilon\right| (s)$ \\
%	\hline
%	\multicolumn{1}{|l|}{Optomechanical system \cite{Pikovski2012_1}} & $10^{-11}$ & $10^5$ & $ 1.61 \times 10^{35}$ \\
%	\multicolumn{1}{|l|}{Bar detector AURIGA \cite{Marin2013_1}} & $1.1 \times 10^{3}$ & $900$ & $1.81 \times 10^{25}$ \\
%	\multicolumn{1}{|l|}{\multirow{4}{*}{Mechanical oscillators \cite{Bawaj2014_1}$ \left\{ \begin{array}{c} ~ \\ ~ \\ ~ \\ ~ \end{array} \right.$}} & $3.3 \times 10^{-5}$ & $5.64 \times 10^3$ & $1.53 \times 10^{31}$ \\
%	\multicolumn{1}{|l|}{}& $7.7 \times 10^{-8}$ & $1.29 \times 10^5$ & $1.26 \times 10^{31}$ \\
%	\multicolumn{1}{|l|}{}& $2 \times 10^{-8}$ & $1.42 \times 10^5$ & $3.99 \times 10^{31}$ \\
%	\multicolumn{1}{|l|}{}& $2 \times 10^{-11}$ & $7.47 \times 10^5$ & $1.44 \times 10^{33}$ \\
%	\multicolumn{1}{|l|}{LIGO detector \cite{Abbott2016}} & 40 & 200 & $1.00 \times 10^{28}$\\
%	\multicolumn{1}{|l|}{GW150914 \cite{Abbott2016}} & $3.2 \times 10^{31}$ & 100 & $0.0502$\\
%	\hline
\end{tabular}
\caption{Period of oscillation of $(\Delta q)^2$ and $(\Delta p)^2$ for several systems, as given by \eqref{eqn:unc_period}. \label{tbl:unc_osc_period}}
\end{center}
\end{table}
We see that in all cases we have a period several orders of magnitude larger than the age of the universe.

As for the uncertainty product, we have
\begin{multline}
	\left(\Delta q\right)^2 \left(\Delta p\right)^2 = 
		\frac{\hbar^2}{4} \left\{1 + 4 |\alpha|^2 \left[1 - e^{- 2 |\alpha|^2 \sin^2 (4 \chi \omega t)}\right]
		+ 2 |\alpha|^4 \left[2 + e^{- |\alpha|^2 \left[3 - 2 e^{8 i \chi \omega t} - e^{- 16 i \chi \omega t} \right] - 8 i \chi \omega t } + \right. \right. \\
		\left. \left.
		+ e^{- |\alpha|^2 \left[3 - 2 e^{-8 i \chi \omega t} - e^{16 i \chi \omega t} \right] + 8 i \chi \omega t } 
		- e^{- 2 |\alpha|^2 \sin^2 \left[8 \chi \omega t\right]} 
		+ e^{- 4 |\alpha|^2 \sin^2 \left[4 \chi \omega t\right]}
		- 4 e^{- 2 |\alpha|^2 \sin^2 \left[4 \chi \omega t\right]} \right] \right\}\\
		- \alpha^4 \frac{\hbar^2}{4} \left\{e^{- |\alpha|^2 \left[1 - e^{16 i \chi \omega t} \right] - 2 i \omega t \left[1 - 12 \chi \right] } - e^{- 2 |\alpha|^2 \left[1 - e^{ 8 i \chi \omega t}\right] - 2 i \omega t \left[1 - 8 \chi \right]} \right\}^2 + \\
		- \alpha^\star {}^4 \frac{\hbar^2}{4} \left\{e^{- |\alpha|^2 \left[1 - e^{- 16 i \chi \omega t} \right] + 2 i \omega t \left[1 - 12 \chi \right] } - e^{- 2 |\alpha|^2 \left[1 - e^{ - 8 i \chi \omega t}\right] + 2 i \omega t \left[1 - \chi \right]} \right\}^2 + \\
		- 8 \frac{\hbar m \omega}{2} \frac{\hbar^2}{4} \left[ (\delta^2-\epsilon) 
			- 2 |\alpha|^2 \epsilon
			+ (\delta^2+\epsilon) \left(\alpha^2 e^{- 2 i \omega t } + \alpha^\star {}^2 e^{2 i \omega t}\right) \right] \gamma^2
		+ 4 i \frac{\hbar^2}{4} \sqrt{\frac{\hbar m \omega}{2}} \delta \left[\alpha e^ {- i \omega t } - \alpha^\star e^{i \omega t}\right] \gamma
\end{multline}
and shows variations similar to those previously analyzed.

 We conclude this section by showing that similar results can be obtained to any order in perturbation theory for a large class of perturbations.
In this case it is convenient to use the Schr\"odinger picture, in which 
each number state component of a coherent state acquires a different phase.
In fact, let us rewrite the Hamiltonian in \eqref{eqn:gen_Hamiltonian} as
\begin{equation}
	H = \hbar \omega \left( \sum_{i=0}^S \kappa_i \tilde{N}^i \right)~,
\end{equation}
where $\kappa_0 = c_0 + \frac{1}{2}$, $\kappa_1 = c_1 + 1$, and $\kappa_i = c_i$ with $i\geq 2$.
A time-evolved coherent state can then be written as
\begin{equation}
	|\alpha,t\rangle = \exp \left[- i \omega t \left( \sum_{i=0}^S \kappa_i \tilde{N}^i \right)\right] |\alpha,0\rangle \propto \sum_{n=0}^\infty \frac{\left(\alpha e^{-i \kappa_1 \omega t}\right)^n}{\sqrt{n!}} \exp \left[- i \omega t \left( \sum_{i=2}^S \kappa_i n^i \right)\right] |n^{(e)}\rangle~,
\end{equation}
where $|n^{(e)}\rangle$ is the perturbed energy eigenstate up to order $\pi$ such that $\tilde{N} |n^{(e)}\rangle = n |n^{(e)}\rangle$.
One can then prove that
\begin{subequations}
\begin{align}
	\langle \alpha,t|q|\alpha,t\rangle \propto & \sum_{n=0}^\infty \frac{\alpha^{2n}}{n!} \cos \left\{ t \left[ \omega \kappa_1 + (\omega_{n+1} - \omega_n) \right] \right\}~,\\
    \langle \alpha,t|p|\alpha,t\rangle \propto & \sum_{n=0}^\infty \frac{\alpha^{2n}}{n!} \sin \left\{ t \left[ \omega \kappa_1 + (\omega_{n+1} - \omega_n) \right] \right\}~,
\end{align}
\end{subequations}
where $\omega_n = \omega \left( \sum_{i=2}^S \kappa_i n^i \right)$.
We then see that the coherent state spreads in phase-space, since each component moves with a different frequency
\begin{equation}
	\Omega_n = \omega \left\{\kappa_1 + \sum_{i=2}^S \kappa_i \sum_{j=1}^i \binom{i}{j} n^{i-j}\right\}~.
\end{equation}
In a frame rotating at the frequency of the ground state, the other states move with a frequency
\begin{equation}
	\Delta \Omega_n = \Omega_n - \Omega_0 = \omega \sum_{i=2}^S \kappa_i \sum_{j=1}^{i-1} \binom{i}{j} n^{i-j}
\end{equation}
We have two cases:
\begin{itemize}
	\item Only one coefficient $\kappa_i$ is different from 0: then all the frequencies are multiple of $\omega_1 = \omega \kappa_i$ and the time interval over which the coherent state spreads and restores is given by
    \begin{equation}
    	T = \frac{2\pi}{|\kappa_i| \omega}~.
    \end{equation}
    
    \item $\kappa_i = \kappa \mu_i$, with $\kappa \in \mathbb{R}$ and $\mu_i \in \mathbb{Q}$: in this case we have
    \begin{equation}
    	T = \frac{2 \pi}{\omega |\kappa|} \frac{\zeta}{\xi}~,
    \end{equation}
    where $\zeta$ is the least common denominator of $\mu_i$'s and $\xi$ is the greatest common divisor of the numerator of $\mu_i$'s.
\end{itemize}
In many cases, this is an over estimation, since when $i$ is a prime number, the actual period is $T' = T/i$, or, if $i$ is a power of a prime number $k$, then one can take $T' = T/k$ as the period.
Finally, notice that the result in \eqref{eqn:unc_period} corresponds to the first case with $\kappa_2=-4\chi$ and that in general the period depends on the HO frequency $\omega$ and on the HO mass $m$ and the GUP parameters included in $\kappa_i$.

\begin{figure}
\center
\begin{subfigure}[t]{0.47\textwidth}
\center
\includegraphics[scale=0.54]{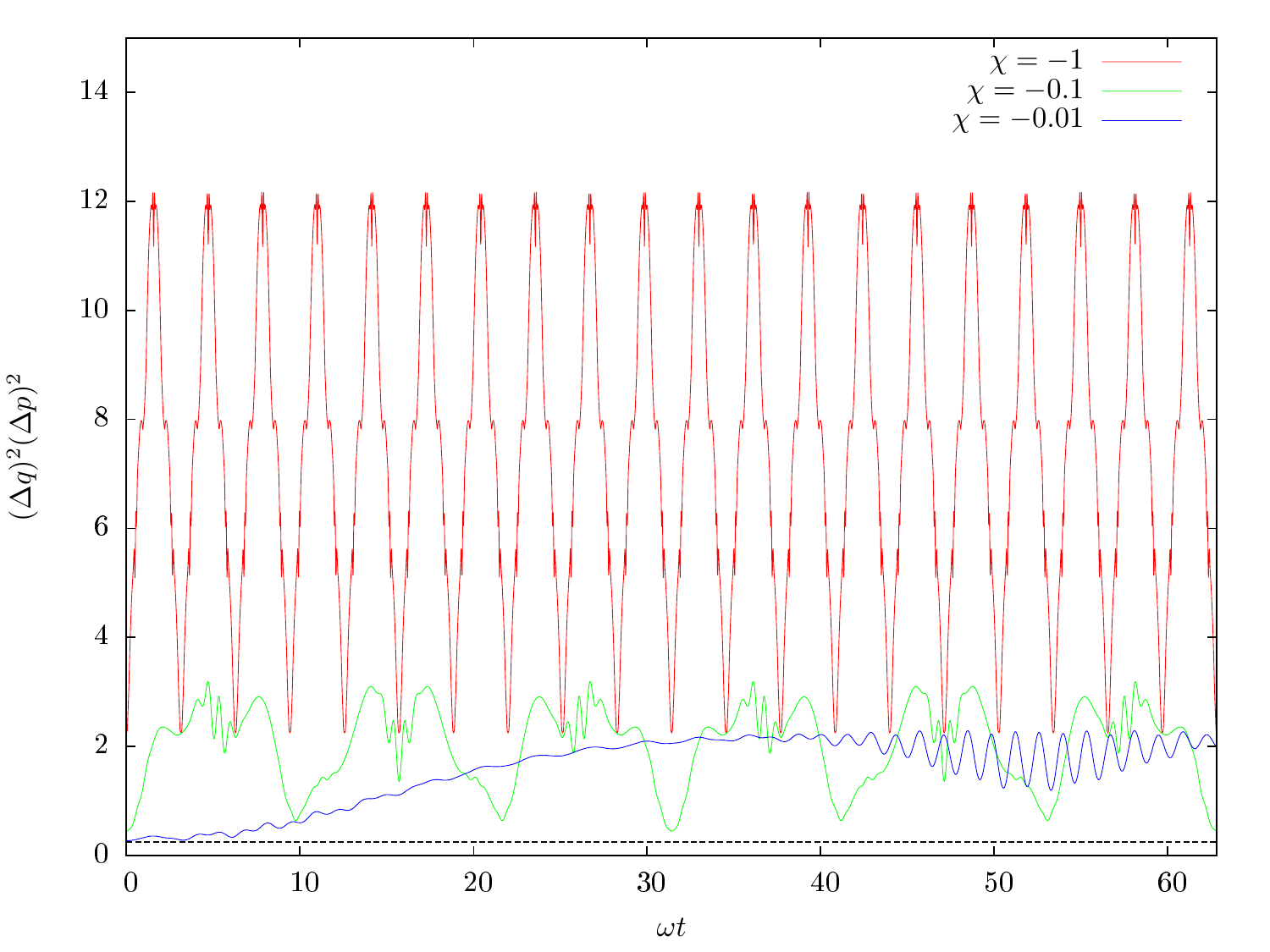}
\caption{The time evolution of the uncertainty product for coherent states is plotted for three values of $\chi$ in a model with $\delta=0$ and $\epsilon=1$ (quadratic GUP).
The black dashed line corresponds to the standard case.
The results are given in natural units.
Notice that each case presents small oscillations superposed to the dominant $\sin^2$ term.
Also, notice that the minimal value of the uncertainty product increases with $|\chi|$.} \label{fig:unc_pro_t_q}
\end{subfigure}
\qquad
\begin{subfigure}[t]{0.47\textwidth}
%\center
%\begingroup
%\fontsize{5}{6} \selectfont
%\include{unc_pro_t_l}
%\endgroup
\includegraphics[scale=0.54]{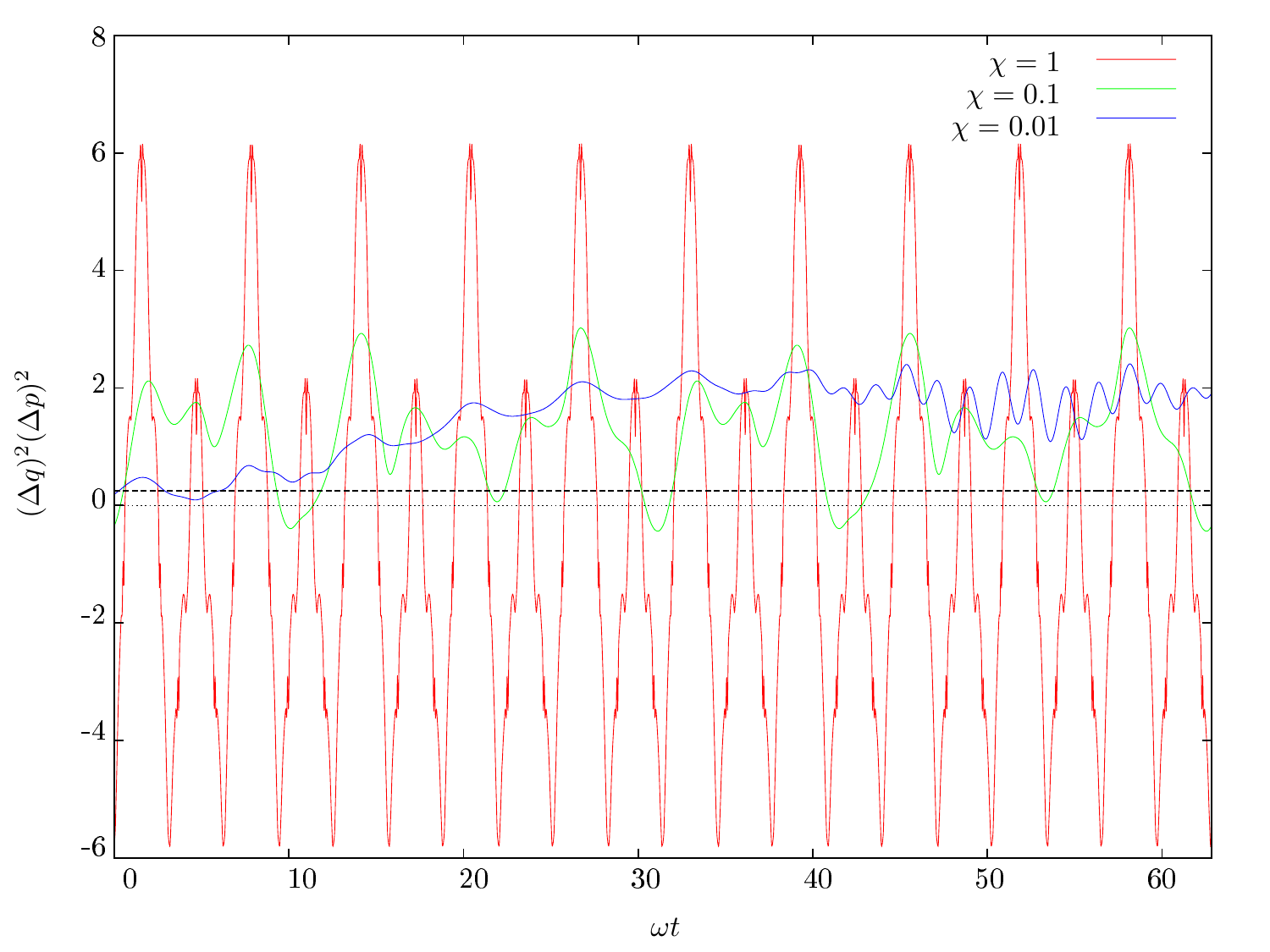}
\caption{Time evolution of the uncertainty product for coherent states for three values of $\chi$ in a model with $\delta=1$ and $\epsilon=0$ (linear GUP).
Similar observations as in the previous figure are valid here.
Here it is evident that a linear model can imply negative uncertainties and uncertainty product.} \label{fig:unc_pro_t_l}
\end{subfigure}
\begin{subfigure}{\textwidth}
\center
%\begingroup
%\fontsize{5}{6}\selectfont
%\include{unc_pro_t_l1}
%\endgroup
\includegraphics[scale=0.54]{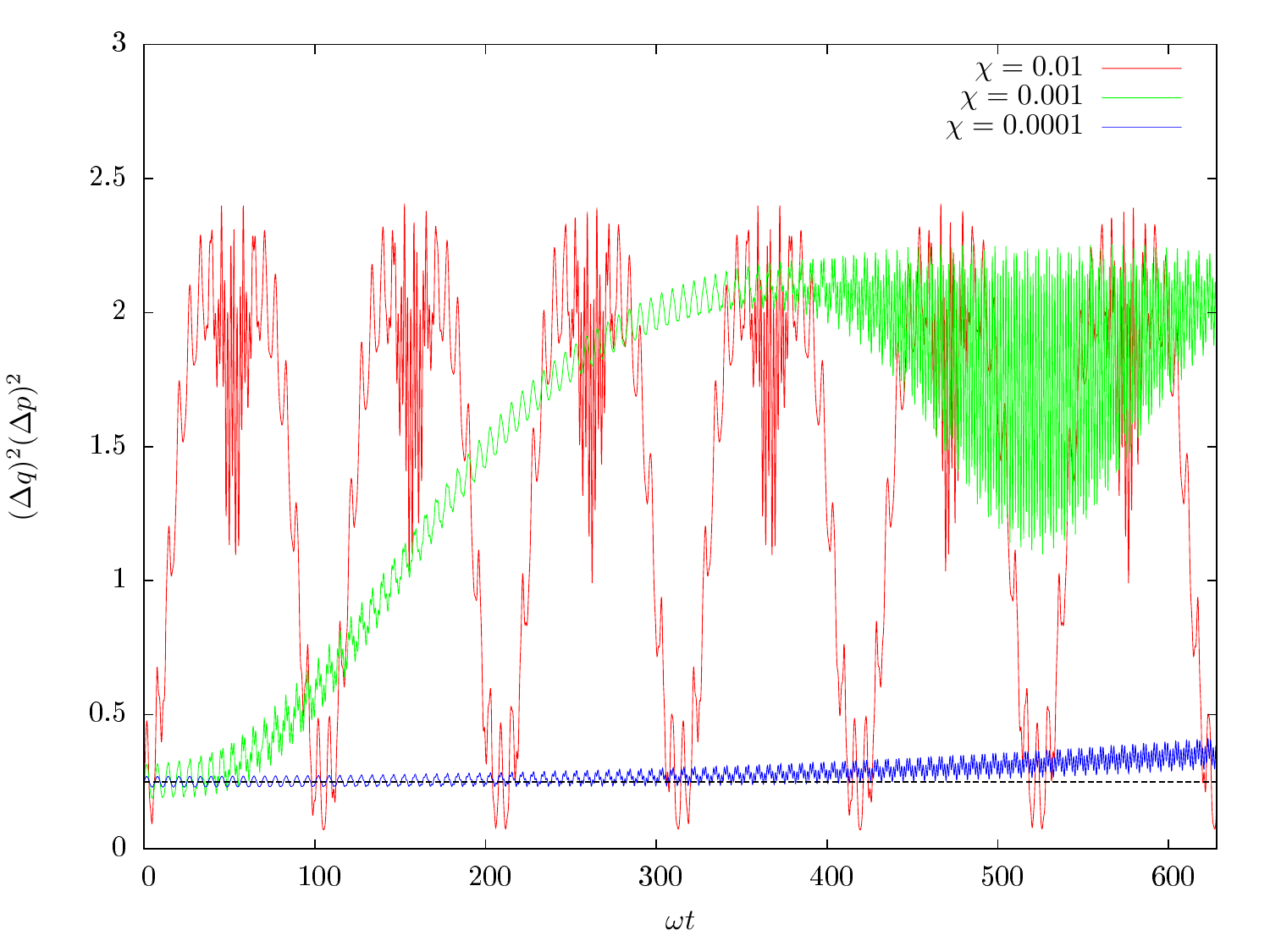}
\caption{Similar case as Figure \ref{fig:unc_pro_t_l}, but with the condition \eqref{eqn:negative_unc_pro} not satisfied. In this case we see that the uncertainty products are positive.} \label{fig:unc_pro_t_l1}
\end{subfigure}
\end{figure}

\section{Squeezed States} \label{sec:squeezed_states}

Next we construct squeezed states with GUP  up to second order in $\gamma$.
Following \cite{Lu1972_1}, we define a new set of operators
\begin{align}
	\tilde{a}_r = & \tilde{a} \cosh r - e^{i\theta} \tilde{a}^\dagger \sinh r~, & \tilde{a}^\dagger_r = \tilde{a}^\dagger \cosh r - e^{- i\theta} \tilde{a} \sinh r~, \label{def:squeezed_ac_operators}
\end{align}
where $r$ is the squeeze parameter and $\theta$ a phase angle.
We can proceed as in the standard theory, defining a squeezed vacuum state through
\begin{equation}
	\tilde{a}_r |0\rangle_r = 0 
\end{equation}
and we find that  $|0\rangle_r$ can be expanded in terms of even numbered states only
\begin{equation}
	|0\rangle_r = \sum_{n=0}^\infty b_{2n} |2n^{(2)}\rangle~,
\end{equation}
with\footnote{Alternatively, to define squeezed states is through the squeeze operator \mbox{$\tilde{\mathcal{S}}(z) = \exp[-\frac{1}{2} ( z \tilde{a}^2 - z^\star \tilde{a}^\dagger {}^2) ]$}, where $z=r e^{i\theta}$.
In this case we have
\begin{align}
	\tilde{a}_z = & \tilde{\mathcal{S}}^\dagger (z) ~ \tilde{a} ~ \tilde{\mathcal{S}}(z) & |0^{(2)}\rangle_z = \tilde{\mathcal{S}}(z) |0^{(2)}\rangle
\end{align}}
\begin{align}
	b_{2n} = & (\tanh r ~ e^{i\theta})^n \sqrt{\frac{(2n-1)!!}{2n!!}} b_0~, & b_0 = & \left[ 1 + \sum_{n=1}^\infty (\tanh r)^{2n} \frac{(2n-1)!!}{2n!!} \right]^{-1/2}~.
\end{align}
Considering the case $\theta = 0$, we can invert the relations (\ref{def:squeezed_ac_operators}) and study the properties of squeezed states using (\ref{eqn:expansions_q&p}), obtaining the following expressions for the uncertainties
\begin{subequations}
\begin{align}
	(\Delta q)^2 = & \frac{\hbar}{2 m \omega} e^{- 2 r} 
	- 4 i (\alpha^\star - \alpha) \delta \frac{\hbar}{2} \sqrt{\frac{\hbar}{2 m \omega}} e^{- 2 r} \gamma 
	+ \epsilon \frac{\hbar^2}{4} (5 + 3 e^{- 4 r} - 2 \alpha^\star {}^2 e^{-2 r} + 16 |\alpha|^2 e^{-2 r} - 2 \alpha^2 e^{-2 r} ) \gamma^2 + \nonumber \\
	& + \delta^2 \frac{\hbar^2}{4} \left[ 11 - 15e^ {- 4 r} + 2 \alpha^2 (2 e^{2 r} - 11 e^{-2 r}) + 8 |\alpha|^2 (e^{2 r} - 2 e^{-2 r}) + 2 \alpha^\star {}^2 (2 e^{2 r} - 11 e^{-2 r}) \right] \gamma^2\\
	(\Delta p)^2 = & \frac{\hbar m \omega}{2} e^{2 r} 
	+ \left(\frac{\hbar m \omega}{2}\right)^2 \left\{ 2 (\alpha^\star {}^2 + \alpha^2) [ 8 e^{-2 r} \delta^2 - 3 e^{2 r} (\delta^2 + \epsilon)] + 8 |\alpha|^2 \delta^2 (4 e^{-2 r} - 3 e^{2 r}) - 3 e^{4 r} (\delta^2 - \epsilon) \right. + \nonumber \\
	& \left. - 3 (3 \delta^2 + \epsilon)  + 8 \delta^2 e^{-4r} \right\} \gamma^2 ~, \\
	(\Delta q)^2 (\Delta p)^2 = & \frac{\hbar^2}{4} \left\{ 1
	- 4 i (\alpha^\star - \alpha) \delta \sqrt{\frac{\hbar m \omega}{2}} \gamma
	+ 4 \frac{\hbar m \omega}{2} \left\{ 2 \epsilon (e^r - \alpha^\star + \alpha) (e^{r} + \alpha^\star - \alpha) + \right. \right. \nonumber \\
	& \qquad \qquad + \delta^2 \left[ \alpha^\star {}^2 (e^{4 r} - 7 + 4 e^{-4r}) + 2 |\alpha|^2 (e^{4 r} - 5 + 4e^{-4 r}) \right. \nonumber \\
	& \qquad \qquad \qquad \left. \left. \left. + \alpha^2 (e^{4 r} - 7 + 4 e^{-4r}) + 2 (e^{2 r} - 3 e^{-2r} + e^{- 6 r}) \right] \right\} \gamma^2 \right\}
\end{align}
\end{subequations}
to order $\gamma^2$. 
As expected, these reduce to quantities in the standard theory for $\gamma \rightarrow 0$.
In this case too, we can compute the theoretical minimal uncertainty through the Schr\"odinger-Robertson relation
\begin{multline}
	[(\Delta q)^2 (\Delta p)^2]_{\mathrm{min}} = \frac{\hbar^2}{4} \left\{ 1 
    	- 4 i (\alpha^\star - \alpha) \delta \sqrt{\frac{\hbar m \omega}{2}} \gamma 
        + 4 \frac{\hbar m \omega}{2} \left\{ 2 \epsilon(e^r - \alpha^\star + \alpha) (e^r + \alpha^\star - \alpha) \gamma^2 + \right. \right. \\
        \left. \phantom{\frac{\hbar m \omega}{2}} \left. - \delta^2 \left[ 3 \alpha^\star {}^2 + 2 |\alpha|^2 + 3 \alpha^2 + 2 e^{-2 r} - (\alpha^\star + \alpha)^2 (e^{2r} - 2 e^{-2r})^2 \right] \right\} \gamma^2 \right\}
\end{multline}
and thus find 
\begin{equation}
	(\Delta q)^2 (\Delta p)^2 - [(\Delta q)^2 (\Delta p)^2]_{\mathrm{min}} = 32 \delta^2 \frac{\hbar^2}{4} \frac{\hbar m \omega}{2} e^{- 2 r} \sinh^2 (2r) \gamma^2 \label{eqn:diff_squeezed}
\end{equation}
for the difference with the minimal uncertainty product.
 Several features are noteworthy. 
First, the above difference is positive for all values of $r$ and   does not depend on $\alpha$.
Second, the difference with the minimal product is of the second order in the GUP parameter.
Finally, it is present only for models with a non-zero linear term; it is not present in the model introduced in \cite{Kempf1995_1}.
Finding evidence for the relation \eqref{eqn:diff_squeezed} could therefore help distinguish the models.

\subsection{Minimal Uncertainties for Squeezed States}

In the standard theory, squeezed states saturate the uncertainty relation allowing for reduced uncertainties, with respect to coherent states, in either position or momentum.
In principle, the uncertainty in position (momentum) can be made arbitrarily small for HUP-saturating states.
However the  GUP  introduces  a minimal uncertainty in position. In this section we study the implications of this fact.

To better observe the features of optimal squeezing in GUP, we consider the particular models examined in \cite{Kempf1995_1}.
This model corresponds to $\delta = 0$ and $\epsilon = 1/4$.
The position uncertainty for this model is
\begin{equation}
	(\Delta q)^2 = \frac{\hbar}{2 m \omega} e^{- 2 r} 
	+ \frac{\hbar^2}{4} (5 + 3 e^{- 4 r} - 2 \alpha^\star {}^2 e^{-2 r} + 16 |\alpha|^2 e^{-2 r} - 2 \alpha^2 e^{-2 r} ) \frac{\gamma^2}{4} ~.
\end{equation}
We obtain a minimal position uncertainty for $e^{-r} = 0 \Rightarrow r = + \infty$.
It is
\begin{equation}
	(\Delta q)^2 = \frac{\hbar^2}{4} \frac{5}{4} \gamma^2  > (\Delta q)^2_{\mathrm{min}} = \hbar^2 \frac{\gamma^2}{4}
\end{equation}
whereas the latter quantity has been previously computed \cite{Kempf1995_1}.
Therefore, the maximally squeezed uncertainty in position is always larger than the minimal uncertainty predicted by the model.

The momentum uncertainty for the same model is
\begin{equation}
	(\Delta p)^2 = \frac{\hbar m \omega}{2} e^{2 r} 
	-\frac{3}{4} \left(\frac{\hbar m \omega}{2}\right)^2 \left\{ 2 (\alpha^\star {}^2 + \alpha^2) e^{2 r} - e^{4 r} + 1 \right\} \gamma^2 ~,
\end{equation}
and is minimized for  $e^{2r} = 0 \Rightarrow r = - \infty$, the minimal uncertainty being
\begin{equation}
	(\Delta p)^2 = - \frac{3}{4} \left(\frac{\hbar m \omega}{2} \right)^2 \gamma^2  
\end{equation}
which is negative. This means that this model cannot describe  infinite squeezing.
Rather, by inverting the reasoning, there must exist a lower value of $r$ for the interval of squeezing in which GUP can be used.
In fact, when we have
\begin{equation}
	e^{2r} = \frac{ - 4 + 3 (\alpha^\star {}^2 + \alpha^2) \hbar m \omega \gamma^2 + \sqrt{ 9\hbar^2 m^2 \omega^2 \gamma^4 + [ 4 - 3 (\alpha^2 + \alpha^\star {}^2) \hbar m \omega \gamma^2 ]^2 }}{3 \hbar m \omega \gamma^2} \simeq \frac{3 \hbar m \omega \gamma^2}{8}~, \label{eqn:max_squeezing_mom_kmm}
\end{equation}
the squeezed uncertainty in momentum vanishes.

We next turn to the consequences for the model in \cite{Ali2011_1}, for which $\delta=1$ and $\epsilon=1$, a case motivated from
doubly special relativity as noted in the introduction.
For simplicity, we consider a squeezed vacuum state, \emph{i.e.} $\alpha=0$.
The position uncertainty in this case is
\begin{equation}
	\left(\Delta q\right)^2 = \frac{\hbar}{2 m \omega}  e^{-2r} + 4 \frac{\hbar^2}{4} (4 - 3 e^{-4r}) \gamma^2~.
\end{equation}
It has a minimum at $e^{-r} = 0 \Rightarrow r = + \infty$, corresponding to
\begin{equation}
	\left(\Delta q\right)^2 = 16 \frac{\hbar^2}{4} \gamma^2 >  \left(\Delta q\right)^2_{\mathrm{min}} = 8 \frac{\hbar^2}{4} \gamma^2
\end{equation}
where the latter quantity is  the minimal length allowed by the model in \cite{Ali2011_1}. 

As for the uncertainty in momentum,  we find
\begin{equation}
	(\Delta p)^2 = \frac{\hbar m \omega}{2} e^{2 r} - 4 \left(\frac{\hbar m \omega}{2}\right)^2 (2 e^{-4r} - 3) \gamma^2 
	\end{equation}
for this model.
Given the presence of both decreasing and increasing exponentials in $r$, there is no minimal value for the momentum uncertainty. Instead
we see that $\Delta p$
  vanishes for
\begin{equation}
	e^{2r_{min}} = 4 \frac{\hbar m \omega}{2} \gamma^2 + \frac{32}{3} \left(\frac{\hbar m \omega}{2} \right)^{2/3} \gamma^{4/3} - 2 \left( \frac{\hbar m \omega}{2} \right)^{1/3} \gamma^{2/3} + \mathcal{O}(\gamma^{8/3})~.
\end{equation}
As with the previous model, this value $r_{min}$ of the squeeze parameter can be interpreted as the smallest physical value of $r$ 
permitted in this model, since smaller values yield $(\Delta p)^2 < 0$.

Finally, in both model we observe similar features: the uncertainties in position have a non-vanishing minimal value larger than the minimal length allowed by the models.
This uncertainties are achieved for infinite squeezing.
On the other hand, minimal uncertainties in momentum, as computed by the GUP models, can be negative, signaling the break down of GUP for high energies.
Therefore, GUP con describe squeezing in momentum uncertainties only up to particular limits for the squeeze parameter.

\section{Applications and Outlook} \label{sec:conclusions}

Various theories of Quantum Gravity predict a modification of the Heisenberg principle and the Heisenberg algebra to include momentum dependent terms.
We consider its implications on the HO from the GUP (\ref{eqn:perturbed_Hamiltonian}) by looking at perturbations from the $\gamma=0$ case  up to second order in $\gamma$.
We were thus able to find normalized eigenstates (new number states) and eigenvalues, and a new set of ladder operators.
We then focused our attention on coherent and squeezed states.

Our most notable results are the new spacing of the energy eigenvalues 
 and the new energy eigenstates  as functions of the linear and quadratic terms $\delta$ and $\epsilon$ in the GUP.
There are three distinct cases.
When $\delta^2 < \epsilon$, the spacing of the energy ladder increases with the number $n$.
For the case $\delta^2 = \epsilon$, the spacing does not depend on $n$, obtaining a regular ladder, as in the standard theory. Finally, for $\delta^2 > \epsilon$ the correction is negative.
Therefore the spacing of the energy ladder decreases with the number $n$ till a maximum number $n_{\mathrm{max}}$, corresponding to the maximum value of the energy.
This simply indicates the GUP breaks down for large enough energies, where Planck scale effects become relevant and a full theory of Quantum Gravity is necessary since higher order terms cannot be neglected.

 We were then able to define a set of modified 
annihilation, creation, and number operators for the perturbed states.
The algebra of these new operators is identical to that of the standard ones.
Furthermore, we show that with their aid, the Hamiltonian can be written in a compact form, from which expectation values, uncertainties, etc. can be computed
quite easily. 
%
%and expectation values and uncertainties result from easier computations than in the case using the operators from standard QM.}

For coherent states, defined as eigenstates of the new annihilation operator, we find that   they still are minimal uncertainty states.
Squeezed states introduce additional interesting features.
The position uncertainty has a lower bound  -- not present in the standard theory -- that depends on the particular GUP model under consideration.  
As for the momentum uncertainty, its maximal squeezing depends on the model and on the choice of the coefficients $\delta$ and $\epsilon$.
That is, for some choices of the coefficients, a lower bound for the squeeze parameter exists, since beyond this limit higher order Planck scale effects cannot be neglected.

Our results   are potentially testable.
For example, direct application to mechanical oscillators can be tested, \emph{e.g.} as in \cite{Pikovski2012_1,Marin2013_1,Bawaj2014_1}.
In fact, mainly motivated by the results in \cite{Kempf1995_1}, many groups already proposed experiments to test them.
So far this class of experiments focused only on the energy spectrum of the HO.
On the other hand, experiments on resonant mass detectors could in principle also test the results concerning coherent and squeezed states \cite{Hollenhorst1979}.
 Actual laboratory based systems, such as those referred above may already be near the threshold of observing such minute corrections.

A different class of tests could be performed 
 for larger systems that can be treated quantum mechanically.
In fact, as we showed in Sec. \ref{sec:HO_GUP} and in particular in (\ref{eqn:detection}), massive systems constitute ideal observational tools to probe GUP effects.
However it is somewhat debatable whether the GUP applies only to fundamental constituents of matter
(\emph{e.g.} \cite{AmelinoCamelia2013_1}), in which case Planck scale effects on mesoscopic and macroscopic objects are small, or if it can be applied to the center of mass of an optomechanical oscillator (and other systems)
(\emph{e.g.} \cite{Pikovski2012_1,Marin2013_1,Bawaj2014_1,Bosso2016_2}), in which case Planck scale effects may be observable with 
current accuracies. For the latter, it will be a
challenge to isolate GUP effects from the rest.  In other words the GUP, being itself a modification of the Heisenberg Uncertainty Principle, should only be applied to systems that exhibit quantum properties.
Indeed, we apply our results in Tables \ref{tbl:detection} and \ref{tbl:unc_osc_period} only for systems that require a quantum description.

It is interesting to comment on the choice of mass scale in \eqref{Mpscale}.
It is generally expected that this is the Planck mass, but on empirical grounds we could replace $M_{\mathrm{Pl}}$ in \eqref{Mpscale} with any mass scale that has yet to be probed by experiment. 
A conservative value would be something just above the LHC scale, \emph{e.g.} $ E_{\mathrm{LHC}} \sim 10$ TeV.
This might come from some hitherto undiscovered intermediate scale, or one deriving from a large 5-th dimension with order TeV Planck energy in five dimensions.
Our estimated corrections  would then be amplified by a factor of $M_{\mathrm{Pl}} c^2 / E_{\mathrm{LHC}} \sim 10^{15}$, or equivalently by a different value of $\gamma_0$ in \eqref{Mpscale}.
Since this ratio appears squared in \eqref{eqn:unc_period}, the most interesting scenario would be the optomechanical oscillator experiments, for which the periods would vary from seconds to hours, potentially rendering the calculated effects observable. 
Alternatively, a scale $E\sim 10^{11}$ GeV would yield oscillation times of seconds for the LIGO detector, again observable at
least in principle.  Depending on the temporal resolution of the detectors, these arguments can set lower bounds to the relevant energy scale for the GUP (see Table \ref{tbl:unc_osc_period}).

Finally, applications to Quantum Optics of the results shown in this paper can lead to important and potentially observable features \cite{Bosso}.
Indeed, starting from a description of the electromagnetic field in terms of amplitude and phase quadratures, it is possible to include the GUP commutation relation to obtain a perturbed Hamiltonian.
Following similar steps and definitions of the present paper, one can then construct a GUP-modified theory of Quantum Optics.
This modification can have important implications on many optical experiments involving squeezed states, in particular future evolution of gravitational wave interferometers.
This is specially relevant as the use of squeezed states have been planned for LIGO interferometers \cite{Dwyer2013_1}.
Therefore, while on one hand further study is still necessary to understand some of the features highlighted in this paper, on the other hand we may be one step closer to experimental evidences of Planck scale Physics.

\section*{Acknowledgements}
This work was supported in part by the Natural Sciences and Engineering Research Council of Canada, Perimeter Institute of Theoretical Physics through their affiliate program, and Quantum Alberta.

%\bibliographystyle{unsrt}
%\bibliography{HO_GUP}

\end{document}